\documentclass{article}
\usepackage{spconf,amsmath,graphicx}
\usepackage{bm} 							
\usepackage{hyperref} 					
\usepackage{url}							
\usepackage{subcaption} 				
\usepackage{amssymb}
\usepackage{mathtools} 					
\usepackage{multirow} 					
\usepackage{hyperref}
\usepackage{booktabs} 					
\usepackage{array}
\usepackage{makecell}

\def\X{{\mathbf{X}}}
\def\H{{\mathbf{H}}}
\def\s{{\mathbf{s}}}

\def\Y{{\mathbf{Y}}}
\def\Rec{{\text{Recog}}}
\def\Syn{{\text{Synth}}}
\def\SpkEnc{{\text{SpkEnc}}}
\def\Dtrg{{\mathbf{D}_\text{trg}}}

\title{S3PRL-VC: OPEN-SOURCE VOICE CONVERSION FRAMEWORK WITH SELF-SUPERVISED SPEECH REPRESENTATIONS}

\name{Wen-Chin Huang$^1$, Shu-wen Yang$^2$, Tomoki Hayashi$^1$, Hung-yi Lee$^2$, Shinji Watanabe$^3$, Tomoki Toda$^1$}
\address{$^1$Nagoya University, Japan\\$^2$National Taiwan University, Taiwan\\$^3$Carnegie Mellon University, USA}

\begin{document}
\ninept
\maketitle

\begin{abstract}
This paper introduces S3PRL-VC, an open-source voice conversion (VC) framework based on the S3PRL toolkit. In the context of recognition-synthesis VC, self-supervised speech representation (S3R) is valuable in its potential to replace the expensive supervised representation adopted by state-of-the-art VC systems. Moreover, we claim that VC is a good probing task for S3R analysis. In this work, we provide a series of in-depth analyses by benchmarking on the two tasks in VCC2020, namely intra-/cross-lingual any-to-one (A2O) VC, as well as an any-to-any (A2A) setting. We also provide comparisons between not only different S3Rs but also top systems in VCC2020 with supervised representations. Systematic objective and subjective evaluation were conducted, and we show that S3R is comparable with VCC2020 top systems in the A2O setting in terms of similarity, and achieves state-of-the-art in S3R-based A2A VC. We believe the extensive analysis, as well as the toolkit itself, contribute to not only the S3R community but also the VC community. The codebase is now open-sourced\footnote{ \url{https://github.com/s3prl/s3prl/tree/master/s3prl/downstream/a2o-vc-vcc2020}}.
\end{abstract}

\begin{keywords}
voice conversion, open-source, self-supervised learning, self-supervised speech representation
\end{keywords}
\section{Introduction}
\label{sec:intro}

\begin{figure}[t]
	\centering
	\includegraphics[width=0.75\columnwidth]{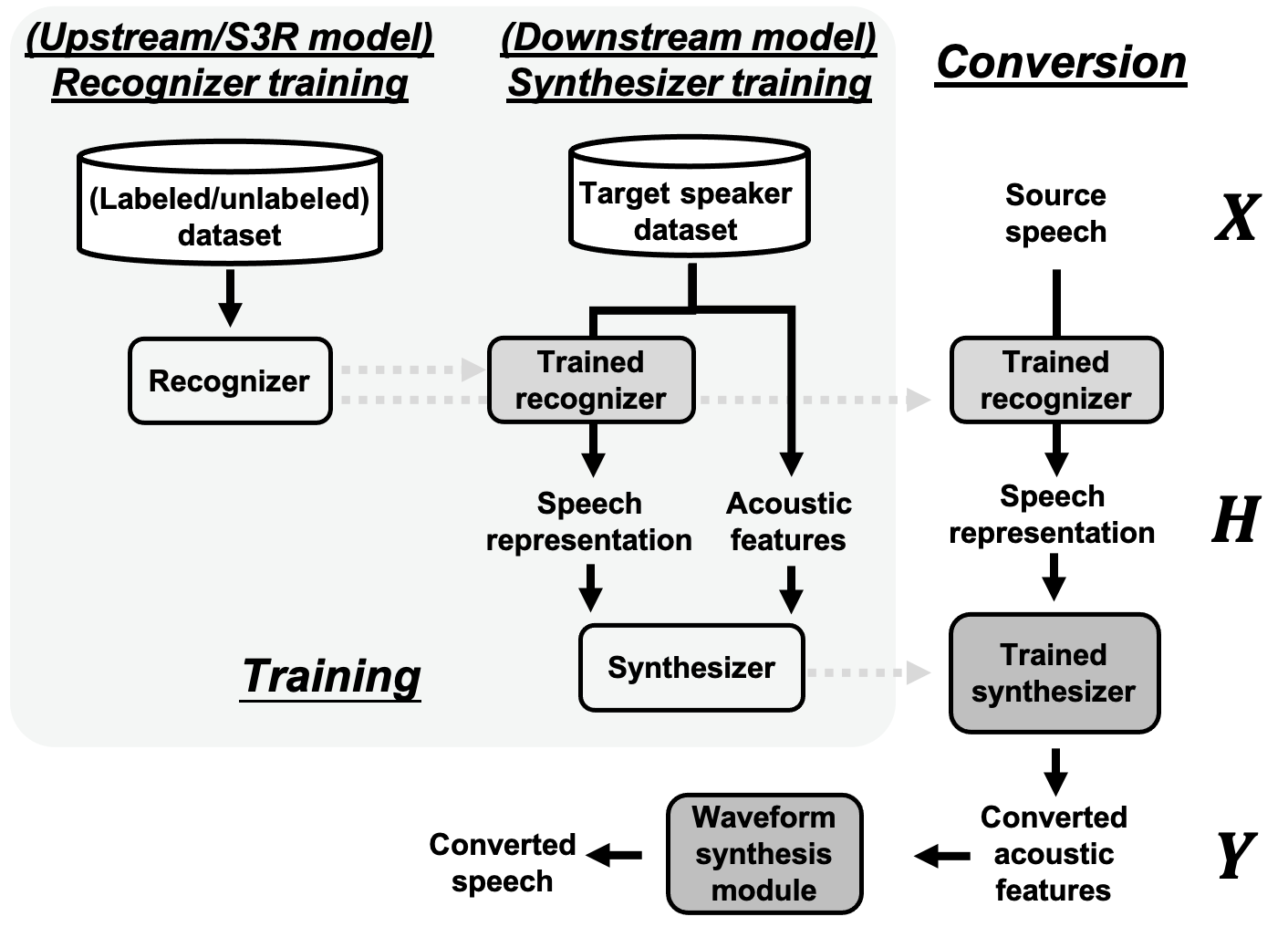} 
	\caption{The training and conversion procedures in any-to-one recognition-synthesis based VC. \label{fig:framework}}
	\vspace{-0.5cm}
\end{figure}

Voice conversion (VC) refers to a technique that converts a certain aspect of speech from a source to that of a target without changing the linguistic content \cite{VC, GMM-VC}. In this work, we focus on speaker conversion, which is the most widely investigated type of VC.
From an information perspective, VC can be performed by first extracting the spoken contents from the source speech, and then synthesizing the converted speech from the extracted contents with the identity of the target speaker.
Such a paradigm is sometimes referred to as recognition-synthesis (rec-syn) based VC, as depicted in Figure~\ref{fig:framework}.
Formally, starting from the source speech $\mathbf{X}$, a recognizer first extracts the spoken contents, $\mathbf{H}$, which is then consumed by the synthesizer to generate the converted speech, $\mathbf{Y}$:
\begin{equation}
    \Y = \Syn(\H), \H=\Rec(\X). \label{eq:formulation}
\end{equation}
In the latest voice conversion challenge 2020 (VCC2020) \cite{vcc2020}, one of the baselines directly concatenated an automatic speech recognition (ASR) model and a text-to-speech (TTS) model \cite{vcc2020-asr-tts}. In addition, several top performing systems also implemented such a framework \cite{vcc2020-task1-top}, showing state-of-the-art performance in terms of both naturalness and similarity.

In rec-syn based VC, an ASR model trained on a labeled dataset is often used to extract the \textit{supervised} spoken content representation, such as text \cite{vcc2020-asr-tts} or phonetic posteriorgram (PPG) \cite{VC-PPG}. The collection of labeled datasets is often costly, especially in a low-resource setting, such as the cross-lingual VC scenario \cite{vcc2020}. Therefore, researchers have resorted to unsupervised or the so-called self-supervised speech representation (S3R) learning paradigm, where a large-scale unlabeled data is used to learn rich, compact speech representations. S3Rs have been applied to any-to-one VC \cite{vqw2v-vc}, many-to-many VC \cite{speech-resynthesis}, any-to-any VC \cite{fragmentvc, s2vc} and cross-lingual VC \cite{prosody-asr-tts}.

In addition to its label-free property, S3R based VC is also attractive in it being a good probing task for S3R analysis. A recently published SUPERB benchmark \cite{superb} dedicates to compare different S3Rs across a range of \textit{discriminative} speech processing tasks, while it remains unclear what representations are optimal for \textit{generation} tasks like VC. For instance, wav2vec 2.0 \cite{wav2vec2} has been shown to be powerful in not only ASR but also speaker and language recognition \cite{wav2vec2-sid-lid}, implying that it encodes rich content, speaker and language information. Based on the discussion on the information perspective of VC, we may hypothesize that a good $\H$ in Eq.~\ref{eq:formulation} should be compact in content but contains little to none speaker information. Based on such an assumption, wav2vec 2.0 may not be an optimal representation for VC.

In this paper, we describe S3PRL-VC, an extension of the S3PRL toolkit and SUPERB.
Our main focus was any-to-one (A2O) VC, where the synthesizer is trained in a target-speaker-dependent fashion. We used the VCC2020 dataset, which allows us to test intra-lingual and cross-lingual settings. We also provide an any-to-any (A2A) extension by using an off-the-shelf d-vector \cite{d-vector} model to encode the unseen speaker information.
We implemented models resembling the top systems in VCC2018 \cite{VC-WNV-adapt} and VCC2020 \cite{vcc2020-task2-top}, which allows us to focus on the comparison.
We conducted a large-scale evaluation, both objectively and subjectively, to compare the performance between not only different S3Rs but also state-of-the-art systems.
S3PRL-VC is a competitive system by yielding (1) a comparable performance with VCC2020 top systems in the A2O setting in terms of similarity, and (2) state-of-the-art performance in S3R-based A2A VC.
Our main contributions are:
\begin{itemize}
	\item Inheriting the property of SUPERB, our S3PRL-VC implementation ensures fast benchmarking but also state-of-the-art performance. Such a fast, easy-to-use property benefits not only S3R researchers but also the VC community.
	\item We present a large-scale comparison of S3Rs from the VC point-of-view, providing new insights and perspectives to analyze the representations. We also compared with top systems in VCC2020 that used PPGs, showing the limitation and competitiveness of S3Rs.
\end{itemize}

\vspace{-0.3cm}
\section{Tasks}

\subsection{General description of VCC2020}

All experiments in this work are benchmarked on the VCC2020 dataset \cite{vcc2020}. There are two tasks in VCC2020, with intra-lingual VC being task 1 and cross-lingual VC being task 2.
The two tasks share the same two English male and female source speakers. The target speakers include two male and two female English speakers for task 1, and one male and one female speaker each of Finnish, German, and Mandarin for task 2.
For each speaker, 70 utterances (roughly five minutes) in their respective languages and contents are provided, and there are 25 test sentences for evaluation.
During conversion, $\X$ (which is in English) is converted as if it was uttered by the target speaker while keeping the linguistic contents unchanged.

\subsection{Intra-lingual and cross-lingual any-to-one VC}

We first consider the two tasks in VCC2020 under the A2O setting. Any-to-one VC aims to convert from any arbitrary speech into that of a predefined target speaker. The training and conversion processes are depicted in Figure~\ref{fig:framework}. The ability to encode $\H$ from any unseen speaker is ensured by the common practice of training S3Rs on a multi-speaker dataset. Using the target speaker dataset, $\Dtrg$, the synthesizer is trained to reconstruct the acoustic feature from $\H$. In the conversion phase, the converted features, $\Y$, are generated following Eq.~\ref{eq:formulation}.
Finally, a waveform synthesizer (ex. neural vocoder) generates the converted waveform.

Any-to-one VC is a good probing task to investigate several characteristics of an upstream S3R model. First, a fundamental requirement of VC is the linguistic consistency, so there is a positive correlation between the VC performance of an S3R model and its ability to faithfully encode $\H$. Second, if an S3R model encodes rich speaker information, then the source speaker information in $\X$ will conflict with the target speaker attributes injected by the synthesizer, which hurts the VC performance. Finally, during the synthesizer training in cross-lingual VC, the S3R model may fail to generalize to $\X$ from a non-English target speaker since most existing S3R models are trained with English datasets only. It is worthwhile to examine the ability of mono-lingual S3R models to transfer to different languages.

\subsection{Intra-lingual any-to-any VC}

We then provide an extension for the A2A scenario, also known as zero-shot VC. A2A VC attempts to convert to a target speaker where $\Dtrg$ is so limited (less than one minute) such that fine-tuning in infeasible. A2A VC models are usually trained on a multi-speaker dataset. Instead of recovering the target speaker information by the synthesizer as in A2O VC, we use speaker embeddings, $s$, extracted by an off-the-shelf speaker encoder, which is pretrained on an automatic speaker verification (ASV) dataset and objective. Such a paradigm is also used in zero-shot TTS \cite{adaptation-verification}. In training, the speaker embedding extracted from the target waveform is used. During conversion, given $\Dtrg$, $\s$ is formed as an average of each embedding from each utterance. We may then rewrite Eq.~\ref{eq:formulation} as:
\begin{equation}
    \vspace{-0.1cm}
    \Y = \Syn(\H, \s), \H=\Rec(\X), \s=\SpkEnc(\Dtrg). \label{eq:a2a-formulation}
\end{equation}

\begin{figure}[t]
	\centering
	\includegraphics[width=\columnwidth]{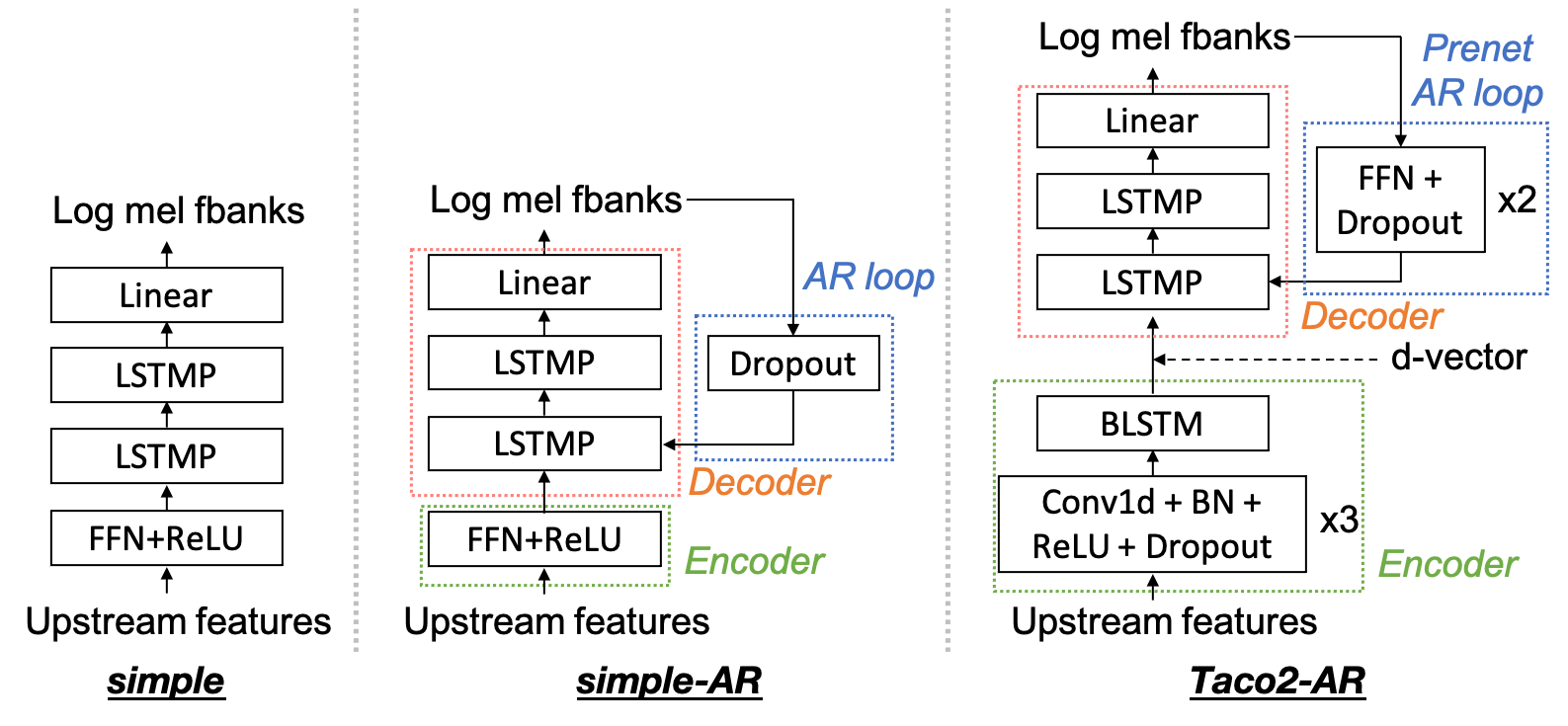} 
	\caption{The models implemented in this work. Left: the simple model. Middle: the simple model with an AR loop. Right: the Tacotron2 model, with extension to an any-to-any model by accepting a d-vector as the speaker embedding. \label{fig:models}}
	\vspace{-0.5cm}
\end{figure}

\begin{table*}[ht!]
\centering

\footnotesize

\caption{Objective evaluation results on different VC settings over various S3Rs. For MCD and WER, the smaller the better; for ASV, the higher the better.}

\label{tab:obj}

\begin{tabular}{|l||r|r|r|r|r|r|r|r|r|r|r|r|r|r|}
\hline
\multirow{3}{*}{Upstream}  & \multicolumn{9}{c|}{Intra-lingual A2O} & \multicolumn{2}{c|}{Cross-lingual A2O} & \multicolumn{3}{c|}{Intra-lingual A2A} \\ \cline{2-15}
& \multicolumn{3}{c|}{Simple} & \multicolumn{3}{c|}{Simple-AR} & \multicolumn{3}{c|}{Taco2-AR} & \multicolumn{2}{c|}{Taco2-AR} & \multicolumn{3}{c|}{Taco2-AR} \\ \cline{2-15}
& MCD  & WER  & ASV  & MCD  & WER  & ASV  & MCD  & WER  & ASV  & WER & \multicolumn{1}{r|}{ASV} & MCD  & WER  & ASV  \\ \hline \hline

mel & 8.41 & 48.5 & 59.00 & 8.92 & 22.7 & 49.75 & 8.47 & 38.3 & 77.25 & 39.0 & 46.67 & 9.49 & 4.2 & 19.50 \\
PPG (TIMIT) & 7.78 & 69.0 & 85.50 & 7.83 & 58.9 & 95.25 & 7.18 & 33.6 & 99.75 & 51.0 & 84.67 & \textbf{8.31} & 12.9 & \textbf{83.50} \\ \hline
PASE+ & 9.29 & 5.0 & 26.75 & 9.52 & 5.7 & 26.00 & 8.66 & 30.6 & 63.20 & 36.3 & 34.67 & 9.85 & 4.2 & 8.00 \\
APC & 8.67 & 8.6 & 48.00 & 8.73 & 7.1 & 41.75 & 8.05 & 27.2 & 87.25 & 33.9 & 52.33 & 9.57 & 3.5 & 23.25 \\
VQ-APC & 8.12 & 10.8 & 81.25 & 8.37 & 7.4 & 60.50 & 7.84 & 22.4 & 94.25 & 28.4 & 68.00 & 9.43 & 4.0 & 22.00 \\
NPC & 7.74 & 39.0 & 92.75 & 8.15 & 21.1 & 76.75 & 7.86 & 30.4 & 94.75 & 37.6 & 59.00 & 9.39 & 4.4 & 21.00 \\
Mockingjay & 8.58 & 31.3 & 51.00 & 8.74 & 9.5 & 47.00 & 8.29 & 35.1 & 79.75 & 39.2 & 46.00 & 9.43 & 5.0 & 25.00 \\
TERA & 8.60 & 11.4 & 46.50 & 8.67 & 6.0 & 42.50 & 8.21 & 25.1 & 83.75 & 29.2 & 49.33 & 9.31 & 5.2 & 18.75 \\
Modified CPC & 8.71 & 9.4 & 40.00 & 8.87 & 7.0 & 30.00 & 8.41 & 26.2 & 71.00 & 35.3 & 32.83 & 9.61 & 4.1 & 10.75 \\
DeCoAR 2.0 & 8.31 & 7.4 & 54.75 & 8.33 & 6.4 & 53.00 & 7.83 & 17.1 & 90.75 & 26.8 & 59.33 & 9.28 & 4.0 & 27.00 \\
wav2vec & 7.45 & 14.0 & \textbf{95.50} & 7.64 & 4.9 & 90.50 & 7.45 & 10.1 & 98.25 & 13.9 & 75.83 & 8.77 & 3.5 & 40.00 \\
vq-wav2vec & \textbf{7.41} & 13.4 & 91.00 & \textbf{7.24} & 11.6 & \textbf{98.75} & \textbf{7.08} & 13.4 & \textbf{100.00} & 21.0 & \textbf{88.83} & 8.47 & 4.2 & 73.25  \\
wav2vec 2.0 Base & 7.80 & 24.7 & 92.75 & 7.77 & 5.0 & 86.50 & 7.50 & 10.5 & 98.00 & 14.9 & 82.17 & 9.03 & 3.2 & 27.00 \\
wav2vec 2.0 Large & 7.64 & 12.5 & 81.75 & 7.67 & 9.0 & 82.75 & 7.63 & 15.8 & 97.25 & 22.7 & 78.00 & 8.99 & 4.1 & 22.25 \\
HuBERT Base & 7.70 & \textbf{5.5} & 89.25 & 7.79 & \textbf{4.7} & 84.25 & 7.47 & \textbf{8.0} & 98.50 & \textbf{13.5} & 82.33 &  9.19 & 3.4 & 23.25 \\
HuBERT Large & 7.54 & 5.6 & 95.00 & 7.54 & 5.6 & 93.00 & 7.22 & 9.0 & 99.25 & 15.9 & 86.50 & 9.13 & \textbf{3.0} & 27.75 \\
\hline
\end{tabular}
\vspace{-0.5cm}
\end{table*}

\vspace{-0.4cm}
\section{Implementation}

\subsection{Recognizer (upstream models)}

Table~\ref{tab:obj} depicts the list of S3Rs we compared in this work, which are the upstream models supported in S3PRL at the date of publication. For a complete list of information (architecture, objective, etc.), refer to \cite{superb}.
All upstreams are trained with English data (mostly Librispeech).
In addition to the S3Rs, two extra upstreams were included: (1) mel-spectrogram, ``mel'', and (2) ``PPG (TIMIT)'', which is trained supervisedly on the TIMIT dataset.

\subsection{Synthesizer model design}
\label{ssec:synthesizer}

Mel-spectrogram was selected as the target acoustic feature. We implemented several models to resemble top systems of past VCCs, as illustrated in Figure~\ref{fig:models}. We avoid expensive model components like attention \cite{transformer} for fast benchmarking.

\noindent\textbf{Simple:} We start from the model used by the top system in VCC2018 \cite{VC-WNV-adapt}. The simple model consists of a single layer feed-forward network (FFN), two long short-term memory layers with projection (LSTMP), and a linear projection layer. 

\noindent\textbf{Simple-AR:} As autoregressive (AR) modeling has been shown to be effective in speech synthesis \cite{ar-rnn-mdn-spss}, we added an AR loop to the simple model. At each time step, the previous output is consumed by the first LSTMP layer. Dropout is essential in the AR loop to avoid exposure bias brought by teacher-forcing \cite{ar-f0-spss, Taco}.

\noindent\textbf{Taco2-AR:} We increase the model complexity by using a model architecture similar to that of Tacotron 2 \cite{Taco2}, which resembles the model used by the top system in VCC2020 \cite{vcc2020-task2-top}. Different from Tacotron 2, the attention module was not used as it was reported to be useless in \cite{vcc2020-task2-top}.

\subsection{Other setups}

\noindent\textbf{Any-to-any settings.} The dataset used to train the A2A VC model is the VCTK dataset \cite{vctk}. For the speaker encoder, we used the d-vector model \cite{d-vector} trained on a mix of datasets, including LibriSpeech, VoxCeleb 1 and 2.

\noindent\textbf{Waveform synthesizer.} We used the HiFi-GAN \cite{hifigan}, a state-of-the-art parallel real-time neural vocoder. For the A2O setup, we mixed the data of all 14 speakers in VCC2020 with the VCTK dataset, while for the A2A setup we used only the VCTK dataset.

\section{Evaluation metrics and protocols}

\subsection{Objective evaluation}

We chose three objective evaluation metrics, all of which measure different aspects of a VC system. Mel cepstrum distortion (MCD) is an intrusive, L2-norm based metric which measures the general performance. Word error rate (WER) measures the intelligibility and the linguistic consistency, and in this work we used a pretrained wav2vec 2.0 model. The accept rate from a pretrained ASV model measures the speaker similarity by calculating the cosine similarity using speaker embeddings. For scenarios like the cross-lingual A2O task where the reference speech is not accessible, we report WER and ASV only since they are non-intrusive.

\subsection{Subjective evaluation}

For the subjective test, we asked listening participants to evaluate two common aspects in VC: naturalness and similarity.
Listeners were asked to evaluate the naturalness on a five-point scale.
For conversion similarity, a natural target speech and a converted speech were presented, and listeners were asked to judge whether the two samples were produced by the same speaker on a four-point scale.

For each system, a total of 80 utterances (5 random $\times$ 16 conversion pairs)  were evaluated. Recordings of the target speakers were also included in the naturalness test and served as the upper bound. 
We used an open-source toolkit \cite{p808-open-source} that implemented the ITU-T Recommendation P.808 \cite{p808} to screen unreliable ratings obtained through the Amazon Mechanical Turk (Mturk). We recruited more than 280 listeners from the United States and had each sample rated by five different participants on average.
Audio samples are available online\footnote{\url{https://bit.ly/3oydaY2}}.

\vspace{-0.2cm}
\section{Evaluation results and discussions}

\subsection{Comparison of different models}

We first investigate the impact of using different synthesizer models described in Section~\ref{ssec:synthesizer} in the intra-lingual A2O setting, as shown in Table~\ref{tab:obj}. First, only by adding the AR loop to the Simple model, most S3Rs benefit from large improvements in WER. With Taco-AR, all S3Rs except PASE+ and modified CPC achieved an ASV accept rate higher 80\%, while all S3Rs suffered from a degradation in WER.
This shows that increasing the model capacity can significantly improve the speaker similarity, while sacrificing the intelligibility.
However, we would like to emphasize that WER is a strict measurement of intelligibility, and human can actually recognize better than machine.
On the other hand, the Taco2-AR model yields the best MCD scores, which, as we will show later, correlates better with subjective naturalness and similarity.
Also, we empirically found the training time of the three models similar.
Based on these reasons, we decided to use the taco2-AR model for the succeeding tasks and comparisons.

\subsection{Results on different tasks}

Next, we compare the results of using S3Rs for different tasks. Looking again at Table~\ref{tab:obj}, we first find S3Rs trained on a mono-lingual corpus can still work well in the cross-lingual setting, demonstrating the ability to transfer across languages.
However, compared with the intra-lingual A2O task, it could be clearly observed that all S3Rs degraded in terms of both the WER and ASV accept rate, which is similar to the findings in \cite{vcc2020-prediction}. Finally, in the intra-lingual A2A setting, all S3Rs yielded WERs much lower than those in the A2O setting, while the MCD values and ASV accept rates were significantly worse. Even the best upstream, vq-wav2vec, yielded only an accept rate of $73.25$. One possible reason is that in the A2A VC setting, modern S3Rs still fail to disentangle content, such that the synthesizer preserves too much speaker information. Another reason may be that a jointly trained speaker encoder \cite{s2vc} is essential for S3R-based VC.

\begin{table}[t]
\centering

\footnotesize

\caption{Comparison with state-of-the-art systems. All upstreams use the Taco2-AR model.}

\label{tab:comparison}

\begin{tabular}{|>{\scriptsize}l||r|r|r|r|c|c|}
\hline
System & MCD & WER & ASV & Naturalness & Similarity \\

\hline \hline
\multicolumn{6}{|c|}{Intra-lingual A2O} \\ \hline
mel & 8.47 & 38.3 & 77.25 & 2.61 $\pm$ 0.11 & 35$\%$ $\pm$ 3$\%$ \\ 
PPG (TIMIT)& 7.18 & 33.6 & 99.75 & 3.32 $\pm$ 0.10 & 58$\%$ $\pm$ 4$\%$ \\
\hline
PASE+ & 8.66 & 30.6 & 63.20 & 2.58 $\pm$ 0.12 & 31$\%$ $\pm$ 3$\%$ \\
APC & 8.05 & 27.2 & 87.25 & 2.92 $\pm$ 0.11 & 43$\%$ $\pm$ 4$\%$ \\
VQ-APC & 7.84 & 22.4 & 94.25 & 3.08 $\pm$ 0.10 & 40$\%$ $\pm$ 4$\%$ \\
NPC & 7.86 & 30.4 & 94.75 & 2.98 $\pm$ 0.11 & 46$\%$ $\pm$ 3$\%$ \\
Mockingjay & 8.29 & 35.1 & 79.75 & 2.81 $\pm$ 0.12 & 42$\%$ $\pm$ 4$\%$ \\
TERA & 8.21 & 25.1 & 83.75 & 2.91 $\pm$ 0.12 & 37$\%$ $\pm$ 4$\%$ \\
Modified CPC & 8.41 & 26.2  & 71.00  & 2.74 $\pm$ 0.11 & 33$\%$ $\pm$ 3$\%$ \\
DeCoAR 2.0  & 7.83  & 17.1  & 90.75  & 3.04 $\pm$ 0.11 & 43$\%$ $\pm$ 4$\%$ \\
wav2vec & 7.45 & 10.1 & 98.25 & 3.40 $\pm$ 0.05 & 52$\%$ $\pm$ 2$\%$ \\
vq-wav2vec & 7.08 & 13.4 & 100.00 & 3.59 $\pm$ 0.10 & 59$\%$ $\pm$ 4$\%$ \\
wav2vec 2.0 B. & 7.50 & 10.5 & 98.00 & 3.36 $\pm$ 0.06 & 51$\%$ $\pm$ 2$\%$ \\
wav2vec 2.0 L. & 7.63 & 15.8 & 97.25 & 3.26 $\pm$ 0.10 & 50$\%$ $\pm$ 4$\%$ \\
HuBERT B. & 7.47 & 8.0 & 98.50 & 3.48 $\pm$ 0.10 & 55$\%$ $\pm$ 4$\%$ \\
HuBERT L. & 7.22 & 9.0 & 99.25 & 3.47 $\pm$ 0.10 & 54$\%$ $\pm$ 4$\%$ \\
\hline
USTC-2018$\dagger$ & -- & 6.5 & 99.00 & 4.20 $\pm$ 0.08 & 55$\%$ $\pm$ 4$\%$ \\
USTC-2020 & 6.98 & 5.4 & 100.00 & 4.41 $\pm$ 0.07 & 82$\%$ $\pm$ 3$\%$ \\
SRCB & 8.90 & 11.5 & 92.00 & 4.16 $\pm$ 0.08 & 68$\%$ $\pm$ 3$\%$ \\
CASIA & 7.13 & 11.0 & 98.25 & 4.25 $\pm$ 0.08 & 61$\%$ $\pm$ 4$\%$ \\
ASR+TTS & 6.48 & 8.2 & 100.00 & 3.84 $\pm$ 0.09 & 75$\%$ $\pm$ 3$\%$ \\
\hline
Target & -- & 0.7 & -- & 4.57 $\pm$ 0.14 & -- \\ 

\hline \hline
\multicolumn{6}{|c|}{Cross-lingual A2O} \\ \hline
PPG (TIMIT)& -- & 51.0 & 84.67 & 2.79 $\pm$ 0.08 & 43$\%$ $\pm$ 3$\%$ \\
\hline
vq-wav2vec & -- & 21.0 & 88.83 & 3.28 $\pm$ 0.08 & 44$\%$ $\pm$ 3$\%$ \\
HuBERT L. & -- & 15.9 & 86.50 & 3.13 $\pm$ 0.08 & 41$\%$ $\pm$ 3$\%$ \\
\hline
USTC-2018 & -- & 5.6 & 97.67 & 4.17 $\pm$ 0.06 & 34$\%$ $\pm$ 3$\%$ \\
USTC-2020 & -- & 7.6 & 96.00 & 4.27 $\pm$ 0.07 & 43$\%$ $\pm$ 3$\%$ \\
SRCB & -- & 8.6 & 78.67 & 4.34 $\pm$ 0.07 & 34$\%$ $\pm$ 3$\%$ \\
CASIA & -- & 10.5 & 91.67 & 4.11 $\pm$ 0.07 & 45$\%$ $\pm$ 3$\%$ \\
ASR+TTS & -- & 34.5 & 67.83 & 2.51 $\pm$ 0.08 & 39$\%$ $\pm$ 3$\%$ \\
\hline
Target & -- & -- & -- & 4.48 $\pm$ 0.12 & -- \\

\hline \hline
\multicolumn{6}{|c|}{Intra-lingual A2A} \\ \hline
PPG (TIMIT) & 8.32 & 12.7 & 84.25 & 3.41 $\pm$ 0.08 & 34$\%$ $\pm$ 4$\%$ \\
\hline
vq-wav2vec & 8.47 & 4.2 & 73.25 & 3.58 $\pm$ 0.09 & 28$\%$ $\pm$ 3$\%$ \\
\hline
S2VC$\dagger$ & -- & 12.4 & 71.50 & 2.90 $\pm$ 0.09 & 29$\%$ $\pm$ 3$\%$ \\
\hline
\multicolumn{6}{l}{\makecell[l]{$\dagger$: Systems generate 16kHz, so MCD is not calculable and direct score\\comparison should be made with caution.}}\\
\end{tabular}
\vspace{-0.5cm}
\end{table}

\begin{table}[t]
\centering
\caption{Linear correlation coefficients between different metrics.}
\label{tab:correlation}

\begin{tabular}{|c||c|c|c|c|c|}
\hline
Metric & MCD & WER & ASV & Nat. & Sim. \\
\hline \hline
MCD & -- & 0.678 & -0.934 & -0.968 & -0.961 \\
WER & -- & -- & -0.640 & -0.808 & -0.587 \\
ASV & -- & -- & -- & 0.910 & 0.911 \\
Nat. & -- & -- & -- & -- & 0.932 \\
Sim. & -- & -- & -- & -- & -- \\
\hline
\end{tabular}
\vspace{-0.5cm}
\end{table}

\subsection{Comparing with top systems using subjective evaluation}

We then compared S3R-based VC models with state-of-the-art systems. \textbf{USTC-2018} \cite{VC-WNV-adapt}, \textbf{USTC-2020} \cite{vcc2020-task1-top, vcc2020-task2-top}\footnote{USTC's systems used text and PPG for the intra-lingual and cross-lingual tasks, respectively.}, \textbf{SRCB} \cite{vcc2020-srcb}, \textbf{CASIA} \cite{vcc2020-casia} were top systems in VCC2020, all of which adopted PPGs, synthesizer pretraining on a multi-speaker dataset, and AR vocoders. Notably, they used thousands of hours of internal data for training. \textbf{ASR+TTS} \cite{vcc2020-asr-tts} was the seq2seq+non-AR vocoder baseline in VCC2020. \textbf{S2VC} \cite{s2vc} is the STOA system for A2A VC. The results are shown in Table~\ref{tab:comparison}. We summarize our observations as follows:
\begin{itemize}
    \item vq-wav2vec outperformed all other upstreams in the subjective test, with a 3.59 naturalness and 59$\%$ similarity in the intra-lingual A2O setting.
    \item In the A2O settings, there was still a naturalness gap between vq-wav2vec and other VCC2020 top systems (3.59 v.s. 4.16-4.25, 3.28 v.s. 4.11-4.34). As for similarity, vq-wav2vec was on par with USTC-2018 and CASIA in the intra-lingual A2O setting, and achieved top in the cross-lingual setting.
    \item In the A2A setting, vq-wav2vec was on par with S2VC in similarity, while being significantly better in naturalness. Our system is therefore the new state-of-the-art in S3R-based A2A VC.
\end{itemize}

\vspace{-0.1cm}
\subsection{Impact of supervision}

Although top systems using PPG greatly outperformed vq-wav2vec in naturalness, they used AR vocoders and the system was trained on large internal datasets, so the impact of supervision is not yet clear. To this end, we compared vq-wav2vec result with ``PPG (TIMIT)'' and the same vocoder. The high WERs and low naturalness scores showed that the PPG was indeed of low quality. Nonetheless, in all three settings, ``PPG (TIMIT)'' can achieve similar or higher similarity scores than vq-wav2vec. This shows that supervision greatly contributes to similarity, especially in difficult settings like A2A VC.
This also shows that the ability of current S3Rs to disentangle speaker information is still limited when compared to PPG, and can be further improved in the future.

\vspace{-0.1cm}
\subsection{Justify the objective metrics with correlation analysis}

Conducting a subjective test whenever a new S3R is developed cannot meet the fast benchmark requirement of SUPERB. Therefore, we examine if the objective measures align well with human perception. Using the intra-lingual A2O results over different upstreams, we calculated pairwise linear correlation coefficients. Results in Table~\ref{tab:correlation} suggested that MCD best aligned with both naturalness and similarity.
Note that in this correlation analysis, we considered systems that used the same decoder and neural vocoder. Since the correlation result is strongly affected by the pool of methods evaluated in a listening test, this good correlation could be observed only in such a homogeneous condition. Nonetheless, this result is still very useful for the benchmarking requirement of SUPERB.

\vspace{-0.1cm}
\section{Conclusions and future work}

We presented S3PRL-VC, an extension of the S3PRL toolkit that applied S3R to VC. We described the model design choice, and covered a variety of tasks. Extensive experiments, both objective and subjective, evaluated the capability of various S3Rs   when applied to different VC scenarios. By comparing S3Rs with supervised presentations like PPG, we showed the competitiveness of S3Rs in certain settings, meanwhile shedding light on improving directions.

We suggest different future directions for readers from different communities. From the VC perspective, it is worthwhile to continue investigating better downstream model design. For instance, in A2A VC, a proper speaker encoder should be used instead of fixed d-vector. Meanwhile, we encourage to use VC as a probing task when designing a new S3R model, considering the challenges to overcome brought by all aspects required in VC.

\noindent{\textbf{Acknowledgements}}
We would like to thank the S3PRL/SUPERB team for the fruitful discussions. This work was partly supported by JSPS KAKENHI Grant Number 21J20920, JST CREST Grant Number JPMJCR19A3, and a project, JPNP20006, commissioned by NEDO, Japan.

\bibliographystyle{IEEEbib}
\bibliography{ref}

\end{document}